\documentclass[twocolumn]{aa}
\usepackage{txfonts}
\usepackage{graphicx,isolatin1}
\def\kms{~km~s$^{-1}$}
\def\cmmt{~cm$^{-3}$}
\def\cmmd{~cm$^{-2}$}
\def\smu{s$^{-1}$}
\def\mum{$\mu$m}
\newcommand{\gsim}{\raisebox{-.4ex}{$\stackrel{>}{\scriptstyle \sim}$}}

\def\le{$\leq$}

\def\HII{H{\sc ii}}

\def\NII{N{\sc ii}}

\def\SIII{S{\sc iii}}
\def\OIII{O{\sc iii}}
\def\OI{O{\sc i}}
\def\CII{C{\sc ii}}
\def\SiI{\mbox{Si {\sc i}}}
\def\SI{\mbox{S {\sc i}}}
\def\SiII{\mbox{Si {\sc ii}}}
\def\FeII{\mbox{Fe {\sc ii}}}
\def\SIII{\mbox{S {\sc iii}}}

\def\le{$\leq$}
\begin{document}
\title{ISO observations of the Galactic center Interstellar Medium:
\thanks{Based on observations with ISO, an ESA project with instruments funded by ESA Member States (especially the PI countries: France, Germany, the Netherlands and the United Kingdom) and with the participation of ISAS and NASA.}
}

\subtitle{neutral gas and dust}

\author{N. J. Rodríguez-Fernández \inst{1}   \fnmsep\thanks{Marie Curie Fellow}
          \and J. Martín-Pintado \inst{2}
%   \fnmsep\thanks{An important part of this work was done at {\it Observatorio
%   Astronomico Nacional}, Alcala de Henares, Spain}
      \and A. Fuente \inst{3}
       \and T. L. Wilson \inst{4}
%        \and S. H\"uttemeister \inst{5}
%  \fnmsep\thanks{Just to show the usage of the elements in the author field}
%   \and P. de Vicente  \inst{3}
   }

   \offprints{N. J. Rodríguez-Fernández}

   \institute{LERMA (UMR 8112), Observatoire de Paris, 61 Av de l'Observatoire,
   F-75014 Paris, France\\
       \email{nemesio.rodriguez@obspm.fr}
       \and
        DAMIR, Instituto de Estructura de la Materia, Consejo Superior de
        Investigaciones Científicas (CSIC), Serrano 121, E-28006 Madrid, Spain
       \and
       Observatorio Astronómico Nacional, Instituto Geográfico Nacional,
       Apdo. 1143, E-28800 Alcalá de Henares, Spain
       \and
       Max-Planck-Institut für Radio Astronomie, Auf dem Hügel 69,
       D53121 Bonn, Germany
%       \and
%       Astronomisches Institut der Ruhr-Universität Bochum,
%       Universitätsstr 150, 44780 Bochum, Germany
     }
   \date{Received ; accepted }

\abstract{ The 500 central pc of the Galaxy (hereafter GC) exhibit a widespread gas component with a kinetic temperature of 100-200~K. The bulk of this gas is not associated to the well-known thermal radio continuum or far infrared sources like Sgr A or Sgr B. How this gas is heated has been a longstanding problem. With the aim of studying the thermal balance of the neutral gas and dust in the GC, we have observed 18 molecular clouds located at projected distances far from thermal continuum sources with the Infrared Space Observatory (ISO). In this paper we present observations of several fine structure lines ([\OI] 63 and 146 \mum, [\CII] 158 \mum, [\SiII] 35 \mum, [\SI] 25 \mum ~ and [\FeII] 26 \mum), which are the main coolants of the gas with kinetic temperatures of several hundred K. We also present the full continuum spectra of the dust between 40 and 190 \mum. All the clouds exhibit a cold dust component with a temperature of $\sim 15$ K. A warmer dust component is also required to fit the spectra. The temperature of this dust component changes between 27 and 42 K from source to source. We have compared the gas and the dust  emission with the predictions from J-type and C-type shocks and photodissociation region (PDRs) models. We conclude that the dust and the fine structure lines observations are best explained by a PDR with a density of  10$^3$ \cmmt ~ and an incident far-ultraviolet field 10$^3$ times higher than the local interstellar radiation field. The fine structure line emission arises in PDRs in the interface between a diffuse ionized gas component and the dense molecular clouds. The [\CII] 158 \mum~ and [\SiII] 35 \mum ~ lines also have an important contribution from the ionized gas component. PDRs can naturally explain the discrepancy between the gas and the dust temperatures. However, these PDRs can only account for 10-30$\%$ of the total H$_2$ column density with a temperature of $\sim 150$ K. We discuss other possible heating mechanisms for the rest the warm molecular gas, such as non-stationary PDRs, X-ray Dominated Regions (XDRs)
or the dissipation of supersonic turbulence.
  }

%\titlerunning{ISO observations of the GC ISM}

\maketitle
%
%________________________________________________________________

%________________________________________________________________

%.........................................

\section{Introduction}

The central 600 pc of the Galaxy (hereafter the Galactic center, GC) contain up to 10$\%$ of the neutral gas of the Milky Way. In spite of the high gas surface density in the GC (several orders of magnitude higher than in the disk of the Galaxy) the star formation rate is 10 times lower than that in the disk and the star formation efficiency is only similar to that in the disk. This is probably due to the particular physical conditions of the GC molecular clouds, which are dense ($\sim 10^4$ \cmmt) but have turbulent linewidths of $\sim$  30 \kms \ and are relatively warm. As shown by the NH$_3$ observations of H\"uttemeister et al. (1993), the GC clouds present a wide range of temperatures from $\sim 15$ K to $\sim 150$ K. They also showed that the warm temperatures are found all over the GC, not only in the vicinity of Sgr A and Sgr B. Regarding the dust, the large scale emission has been studied  in the photometric  far-infrared (between 40 and 250 \mum) survey by Odenwald and Fazio (1984). Their map is relatively similar to the radio continuum maps, showing some discrete sources over an extended emission. They derived a dust temperature of $\leq 30$ K. Recently, Pierce-Price et al. (2000) have derived an homogeneous  dust temperature of 20 K in the GC from 450 and 850 \mum~ data. Clearly, the dust temperature in the GC molecular clouds is considerably lower than the gas temperature. The thermal structure of the GC clouds is difficult to explain but it is required a mechanism that heats selectively the gas and acts over large regions. Radiative heating mechanisms are usually ruled out with the argument that most of the gas does not seem to be associated with thermal continuum sources and that the dust temperature is much lower than that of the gas. Several alternative heating mechanisms have been discussed in the literature as cosmic-rays or magnetic heating (Morris et al. 1983, Gusten et al. 1981, H\"uttemeister et  al. 1993). However, the most promising heating mechanism is shocks or the dissipation of supersonic turbulence (Wilson et al. 1982, Güsten et al. 1985).

We have selected a sample of 18 sources located in various parts of  the GC with the aim of having further insight in the physical properties of the GC interstellar medium. The sources were selected as molecular peaks in CS and SiO large scale maps (Bally et al. 1987, Martín-Pintado et al. 1997, 2000).
They are located far from thermal radio continuum or far-infrared sources in order to study the thermal balance of the interstellar gas in the sources where the origin of the high temperatures is more difficult to explain. 
The typical sizes of the clouds are 5-10 pc. Rodriguez-Fernandez et al. (2001a) have measured H$_2$ densities of (5-20)\,10$^3$ \cmmt ~ and column densities of $\sim (1-6)\,10^{22}$ \cmmd ~ in these sources. The mass of the clouds ranges between 10$^4$ and 10$^5$ M$_\odot$. Therefore, they are representative of the GC molecular clouds (see Miyazaki \& Tsuboi 2000 for a statistical study of the GC clouds population).

We have observed these clouds with the LWS and SWS instruments onboard the Infrared Space Observatory (ISO) since they are perfectly suited to study the thermal balance of the GC clouds. For instance, H$_2$ pure-rotational lines observations have allowed to measure the total amount of warm gas in the GC clouds (Rodríguez-Fernández et al. 2000, 2001a). On average, the column density of warm H$_2$ is $\sim 30 \%$ of the H$_2$ column density traced by CO. However, in some sources it can be much higher (up to 100~$\%$). The warm H$_2$ is mainly gas with a temperature of 150 K. On the other hand, fine structure lines of different ions allow us  to  probe the low density ionized gas and to estimate the properties of the ionizing radiation (Rodríguez-Fernández and Martín-Pintado 2004). The whole GC region is permeated by radiation with an effective temperature of $\sim 35000$ K. ISO also gives us the possibility of studying the dust continuum spectrum around the emission peak (between 40 and 190 \mum) and fine structure lines as [\OI] 63 \mum, [\CII] 158 \mum ~ or [\SiII] 34 \mum, which are the main coolants of the neutral gas with kinetic temperatures of a few hundred K. This  paper is organized as follows: Section 2 presents the observations and explains the data reduction procedures. In Sects. 3 and 4 we report on the results regarding the dust continuum and the fine structure lines. Section 5 revisits the H$_2$ emission and Sect. 6 is devoted to a discussion of the implications of those results. We summarize our conclusions in Sect. 7.

\begin{figure*}[tbh!]
\caption{Position of the observed sources (black diamonds) on the 20 \mum \ image by the MSX satellite (grey levels) and the C$^{18}$O map by Dahmen et al. (1997, contours). M+3.06+0.34 and M+2.99-0.06 (located in the Clump 2 complex) are not shown in this figure}
\label{fig_lws01_cont}
\end{figure*}

\begin{figure*}[tbh!]
\includegraphics[width=18cm]{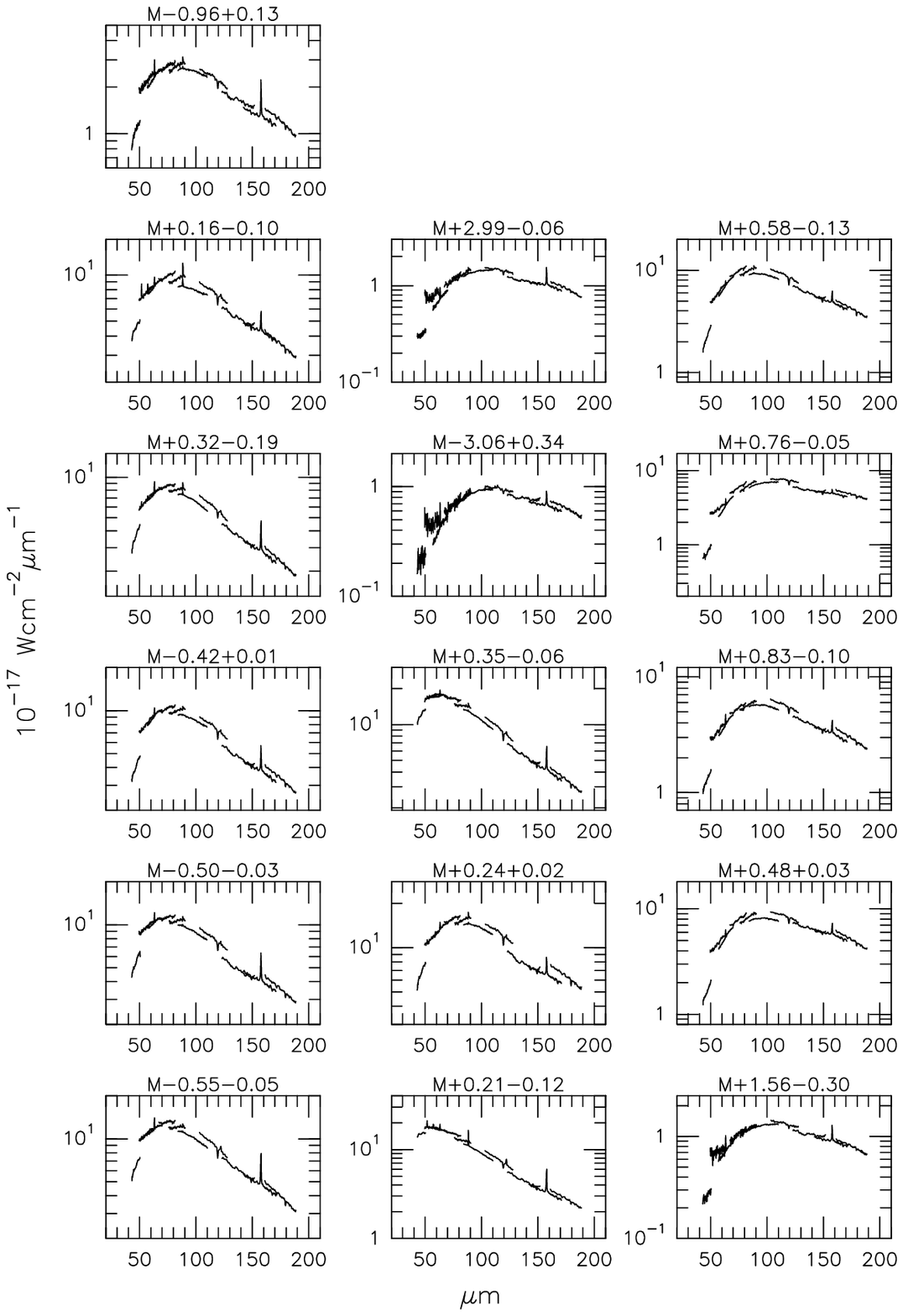}
\caption{LWS01  continuum spectra}
\label{fig_lws01_cont}
\end{figure*}

%__________________________________________________________________
\section{Observations and data reduction}

\begin{table*}[tb]
\caption[]{Integrated flux densities of the lines observed in the SWS02 and LWS01 modes. Numbers in parentheses are the errors of the last significant digit}
\label{tab_grating}
\begin{tabular}{lllllll}
\hline
\noalign{\smallskip}
Line      & SI & FeII & SiII & \OI     &\OI          &\CII   \\
$\lambda (\mu\mathrm{m})$ & 25.24 & 25.98& 34.81 &63 \mum &145 \mum  &158 \mum  \\
Aper. ($^{"}\times^{"}$)  & 14$\times$27 & 14$\times$27 & 20$\times$33 & $86\times86$ & $70\times70$ & $68\times68$\\
Res. (\kms) & 230 & 230 & 207 & 1400 & 1240  & 1140  \\
Units  W\,cm$^{-2}$   &10$^{-20}$ &10$^{-20}$ &10$^{-20}$ & 10$^{-19}$ & 10$^{-19}$ & 10$^{-19}$  \\
\noalign{\smallskip}
\hline
\noalign{\smallskip}
M--0.96+0.13 &\le0.79 &\le0.55& 38(2)  & 24.0(11) & \le 2.4   & 65(3)   \\
M--0.55--0.05&\le0.90 &\le1.6 & 158(5) & 77(4)    & \le14.0   & 242 (7) \\
M--0.50--0.03&\le0.59 &1.8(4) &116(5)  &60(2)     & \le 8.0   & 164(7)  \\
M--0.42+0.01 & \le0.56&1.8(2) &148(4)  & 45(3)    & \le 7.2   & 145(6)  \\
M--0.32--0.19&\le 0.80&1.0(4) &106(4)  & 59.9(9)  &\le  7.2   & 142(6)  \\
M--0.15--0.07& \le0.64& 3.9(3)& 196(5) & ...      & ...       & ...     \\
M+0.16--0.10 & \le0.65&\le1.1 & 121(3) & 39(3)    & \le 11.0  & 114(8)  \\
M+0.21--0.12 &\le 0.68&13.1(6)& 580(20)& 83(6)    & \le 10.0  & 183(6)  \\
M+0.24+0.02  &\le 0.62&1.3(3) & 146(4) &53(2)     & \le 18.0  & 140(7)  \\
M+0.35--0.06 &\le 1.0 &1.7(4) & 154(3) & 67.1(9)  & \le 8.6   & 168(8)  \\
M+0.48+0.03  &\le 0.89&1.0(2) &33.6(10)& 31.6(9)  & \le 9.9   & 107(7)  \\
M+0.58--0.13 &\le 0.77&\le0.91& 42(2)  &27.8(11)  &\le  9.4   & 100(5)  \\
M+0.76--0.05 &\le 0.58&\le0.51&23.9(10)& 24.9(13) &\le  8.5   &51(7)    \\
M+0.83--0.10 &\le 1.2 &\le0.66&20(2)   & 27.4(13) & \le 4.7   & 61(4)   \\
M+0.94--0.36 &\le 0.6 &\le0.67&5.5(13) & ...      & ...       & ...     \\
M+1.56--0.30 & \le0.71&\le0.49& 5(2)   & 10.1(6)  &\le  2.0   & 25.1(6) \\
M+2.99--0.06 &\le 0.89&\le0.51& 17.0(8)& 8.9(6)   &\le 1.7    & 36(2)   \\
M+3.06+0.34  & \le0.70&\le0.49&  2.9(6)&9.6(8)    & \le 1.5   & 18.8(13)\\
\noalign{\smallskip}
\hline
\end{tabular}
\end{table*}

\begin{table*}[tb]
\caption[]{Integrated flux densities (in units of 10$^{-18}$ W\,cm$^{-2}$), widths and peak velocities (in \kms) of the lines observed in the LWS04 mode. Numbers in parentheses are the errors of the last significant digit}
\label{tab_lws04}
{\footnotesize
\begin{tabular}{llllllllllll}
\hline
\noalign{\smallskip}
Source&\multicolumn{3}{c}{\CII~157.74~\mum}&
\multicolumn{3}{c}{\OI~63.18~\mum}&
\multicolumn{3}{c}{\OI~145.52~\mum}\\
& Flux& $\Delta v$ & $v_{hel}$&
Flux  & $\Delta v$ & $v_{hel}$&
Flux  & $\Delta v$ & $v_{hel}$ \\
\noalign{\smallskip}
\hline
\noalign{\smallskip}
M-0.96+0.13&1.9(2)  & 61 & 0 &\le1.4  & (100) & & \le0.3  & (100)   \\
 ~         &2.6     &150 & -100&&&&&& \\
M-0.55-0.05 &5.0(7)  & 83 & -100& 2.4(6)  &100 &-110 &\le0.5   &(200) & ...    \\
~       &    8.2(6)&120&20& 1.1(6) & 71 &26 & & &  \\
M-0.50+0.03& 12.5(1000)   & 110 &-140 &4.1(150)  & 110 & -160 & \le2  &(200) &...   \\
~        & 7.2(1000)     & 85  &-28 & 0.8(140)  & 40& -48 &&&\\
M-0.42-0.01&5.1  &87  & -146 & 3.4(7)  &150&-130 &\le0.4  & (200)&     \\
 ~  &11.2 &135 & -30&&&&&&\\
M-0.32-0.19&5.7(6)  & 60  & 0 &  1.0  & 60  &20 & \le 1.3 & (110)      \\
~         &10      & 200 & -65 & 2.8&170&-65&&&\\
M-0.15-0.07 & ...    &  ...  & ... & ... & ... & ...   & ... & ... & ...   \\
M+0.16-0.10& 10.2(8)  & 122  & 20  & 3.2(4)   & 130  & 39 & \le 0.6  &(150)&  \\
M+0.21-0.12&20.8(9)  & 117 & 16.9& 8.0(10)  & 93  & 36  & 0.65(14) & 42& 20    \\
M+0.24+0.02& 18.3(13)  & 106 &  14 & 5.1(6)  &108  &33 & 2.1(4)  &104 & 20        \\
M+0.35-0.05&19.9(14) & 87    & 27 & 7.1(9)   & 75    & 36 & 1.1(2)   & 73    & 36   \\
M+0.48+0.03& 11.2(5)   & 88   & 30  & 2.0(4)   & 50    & 38 & 0.8(2)   &  95  &  43   \\
M+0.58-0.13&12.8(7)   & 120  & 39 & 4.9(5)   & 109    & 51& 0.55(7) &  61  &  67  \\
M+0.76-0.05& 2.5(200)  & 59 & -14   &\le1.8  & (140) & & \le0.5  & (140)&    \\
             & 3.6(200)  & 95 & 68 &&&&&&  \\
M+0.83-0.10&4(4)   & 78   & -8 & \le 2    & (140) &   & \le 0.6  & (140)     \\
              &3.9(4) & 72   & 90  &&&&&&\\
M+0.94-0.36 &  & ... & ... & ...   & ... & ... & ...  & ... &  ...   \\
M+1.56-0.30&\le1.2   & (150) &  &\le0.8  & (150) & & \le0.3  & (150)   \\
M+2.99-0.06&2.4(2)  & 74 & -14 &\le0.8  & (100) & & \le0.3  & (100)   \\
M+3.06+0.34&\le0.8  & (100) &   &\le0.8  & (100) & & \le0.6  & (100)  \\
\noalign{\smallskip}
\hline
\end{tabular}
}
\end{table*}

\begin{figure}[tbh]
\includegraphics[width=8cm]{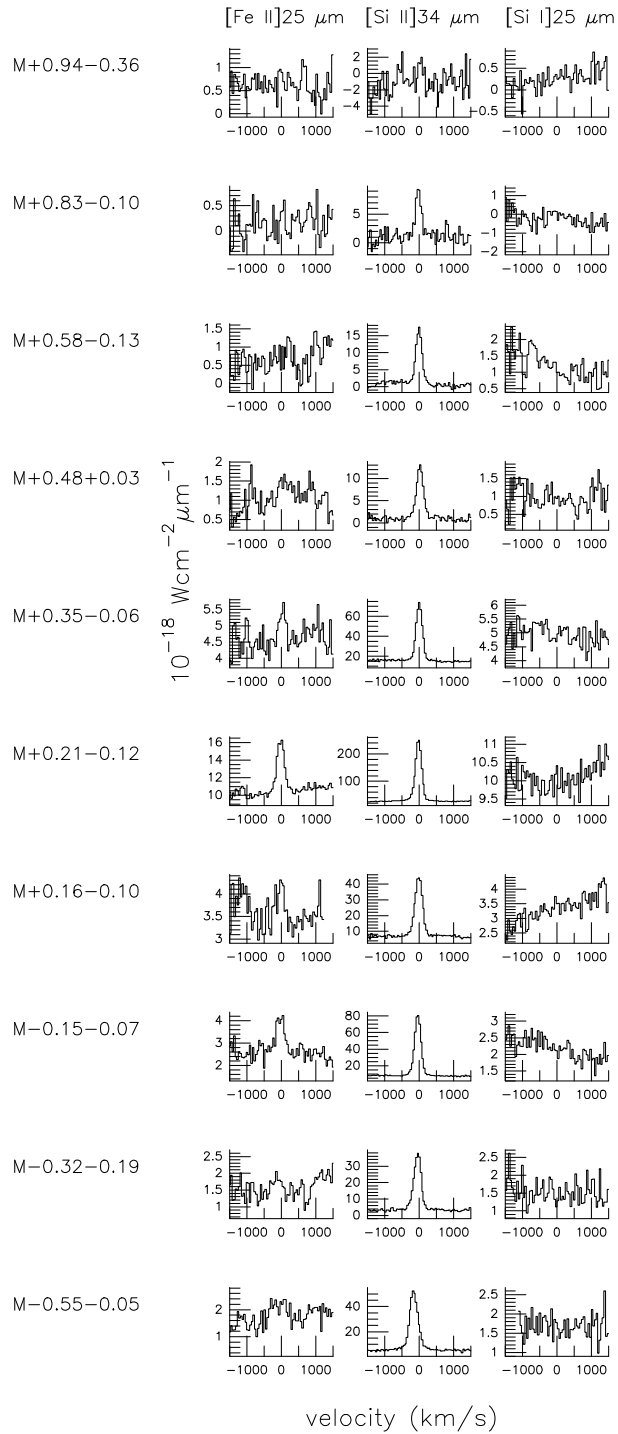}
\caption{Sample of the lines observed in the SWS02 mode}
\label{fig_sws02}
\end{figure}

\begin{figure}[tbh]
\includegraphics[width=8cm]{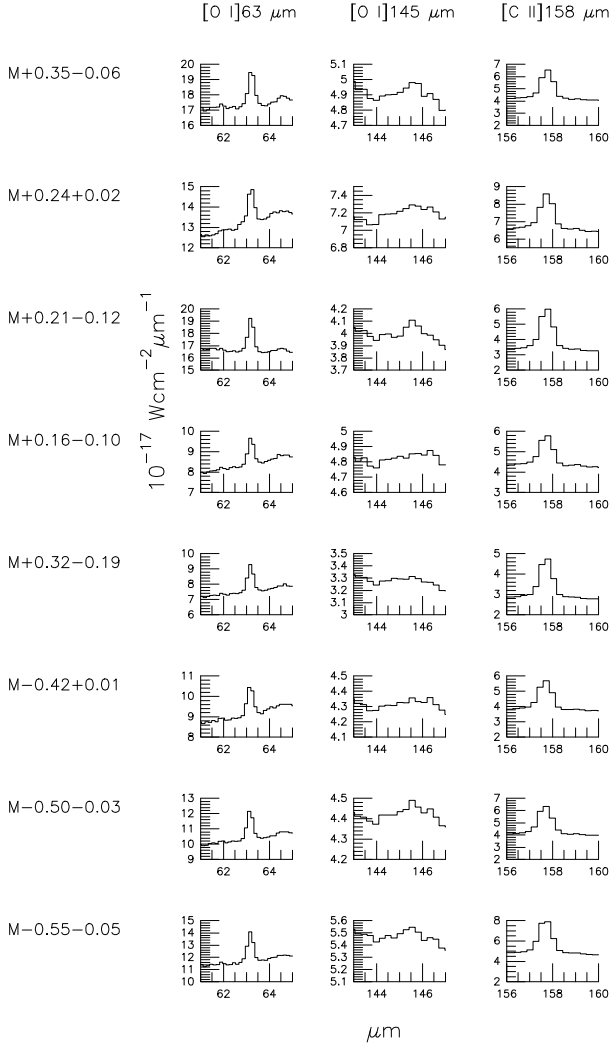}
\caption{Sample of LWS01 spectra of selected lines}
\label{fig_lws01}
\end{figure}

\begin{figure}[tbh]
\includegraphics[width=8cm]{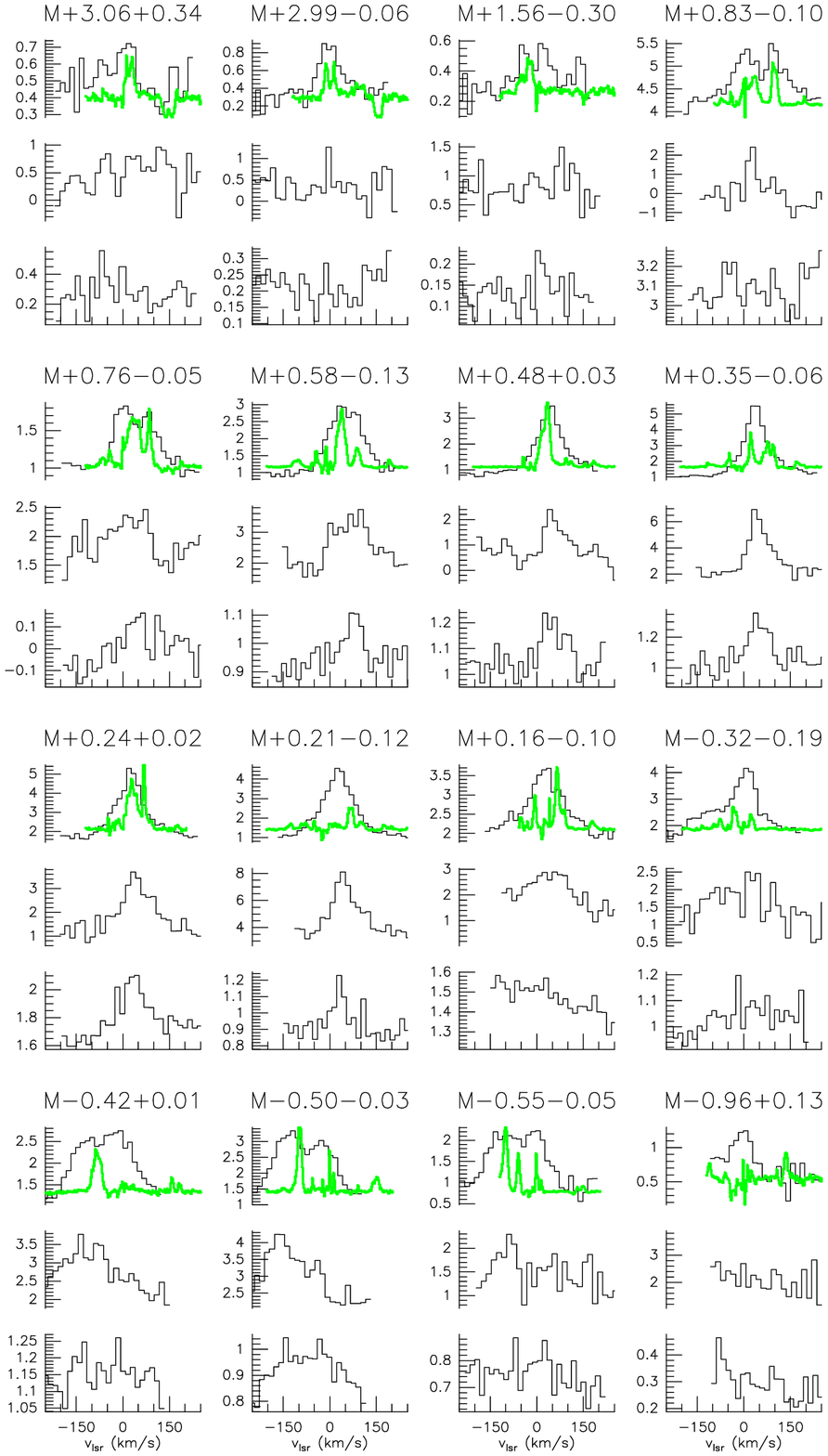}
\caption{LWS04 spectra. For each source the lower spectrum is the [\OI] 145 \mum ~line, that in the middle the [\OI] 63 \mum line and the upper one the [\CII] 158 \mum ~ line. In addition the upper panels also show $^{13}$CO spectra (gray lines).  All the LWS04 spectra are represented in units of $10^{17}$ W\,cm$^{-2}\mu$m$^{-1}$. The $^{13}$CO spectra are represented in arbitrary  units.}
\label{FPvsCO_all.eps}
\end{figure}

We have taken LWS continuous spectra (43-197 \mum) in grating mode (LWS01). The spectral resolution of this mode is 0.29 \mum~ for the first five detectors (43-93 \mum) and 0.6 \mum \, for the rest (84-197 \mum). The aperture of the LWS is $\sim 80^{''} \times 80^{''}$. At the resolution of the LWS01 mode a number of fine structure lines are detected. Those arising exclusively from \HII~ regions are presented in Rodríguez-Fernández and Martín-Pintado (2004). Here we present the 63 and 145 \mum~ lines of \OI~  and the 158 \mum ~ line of \CII. We have also taken [\SiI] 25.2 \mum, [\FeII] 25.9 \mum ~ and [\SiII] 34.8 \mum ~ spectra with the {\it Short Wavelength Spectrometer} in grating mode (SWS02 mode).  The line wavelengths as well as the telescope aperture and the velocity resolution at those wavelengths are shown in Table \ref{tab_grating}. In addition, the [\OI] 63 and 145 \mum ~ and the [\CII] 158 \mum ~ lines have also been observed with the LWS Fabry-Perot (LWS04 mode). The velocity resolution in this mode is $\sim 30$ \kms~ and the wavelength calibration uncertainties expressed in units of velocity are 5 \kms ~ for the [\OI] 63 and [\CII] 158 \mum~ and 15 \kms ~ for the [\OI]145 \mum.

All of  the observations were processed using the off-line processing (OLP) software version 10. Further reduction was done with ISAP. The data reduction  consists basically in dropping bad data points, shifting  the different scans taken with each detector to a common level and averaging all the data for each detector. In addition, with ISAP we have defringed the LWS01 spectra to get rid of the well-known interference pattern. In the overlapping regions of the LWS detectors we have determined that the discrepancies in the fluxes measured by the different detectors are lower than 15$\%$. We  have not tried to shift the different detectors to obtain a smooth continuum spectrum since there is no reliable way of doing this. The absolute flux calibration uncertainties of the LWS01 mode are lower than $\sim 20\%$. The dust continuum spectra are shown in Fig.\ref{fig_lws01_cont}.

In Figs. \ref{fig_sws02} and \ref{fig_lws01} we show a sample of the lines observed in the SWS02 and LWS01 modes. Table \ref{tab_grating} contains the line fluxes as derived by fitting Gaussian curves to the lines. The absolute flux calibration uncertainties of the SWS02 observing mode are less than  30 $\%$.  Figure \ref{FPvsCO_all.eps} shows a sample of the spectra obtained in the LWS04 mode. The results of fitting Gaussian curves to the lines (fluxes, linewidths and the velocities of the center of the lines) are listed in Table \ref{tab_lws04}. The line flux calibration uncertainties of this mode are $\sim 20 \%$. At the resolution of the LWS04 mode all the lines have been  resolved. Most of these have very broad lines, with linewidths up to $\sim 150$ \kms. The central velocities and the profiles of the different lines detected in a given  source are in agreement with each other. The LWS04 and LWS01 lines fluxes are  in agreement within the  calibration uncertainties.

%....................................................
\section{Dust}

We have fitted the dust continuum  spectra  using the Levenberg-Marquardt method. It is not  possible to obtain a good fit to the spectra with just one gray body. Thus, we have used a model with two gray bodies in which, $I_\lambda$, the total intensity radiated at wavelength $\lambda$ is defined as:

\begin{equation}
  I_\lambda=
(1-e^{-\tau_\lambda^{w}})\;B_\lambda(T_{w})\;\Omega_{w}
+ (1-e^{-\tau_\lambda^{c}})\;B_\lambda(T_{c})\;\Omega_{c}
\label{eq-flujoBB}
\end{equation}

where subindex $c$ and $w$ refer to a cold and a warm component, respectively.
$B_\lambda(T)$ is the Planck function at temperature $T$, $\Omega$ is the solid angle of the emission, and $\tau_\lambda$ the dust opacity at wavelength $\lambda$. Since the dust emission in the GC is extended we have used $\Omega_{w}=\Omega_{c}=\Omega_{LWS}$ ($\Omega_{LWS}$ represents the LWS beam). To obtain the dust opacity at a given $\lambda$,  we have  extrapolated the 30 \mum ~ opacity ($\tau_{30}$) using a power law with exponent $\beta$:

\begin{equation}
\tau_\lambda \; = \; \tau_{30} \;\left(30/\lambda_{\mu m} \right)^{\beta}
\label{eq-opacidad}
\end{equation}

\begin{figure}[h]
\centering
\includegraphics[angle=-90,width=7cm]{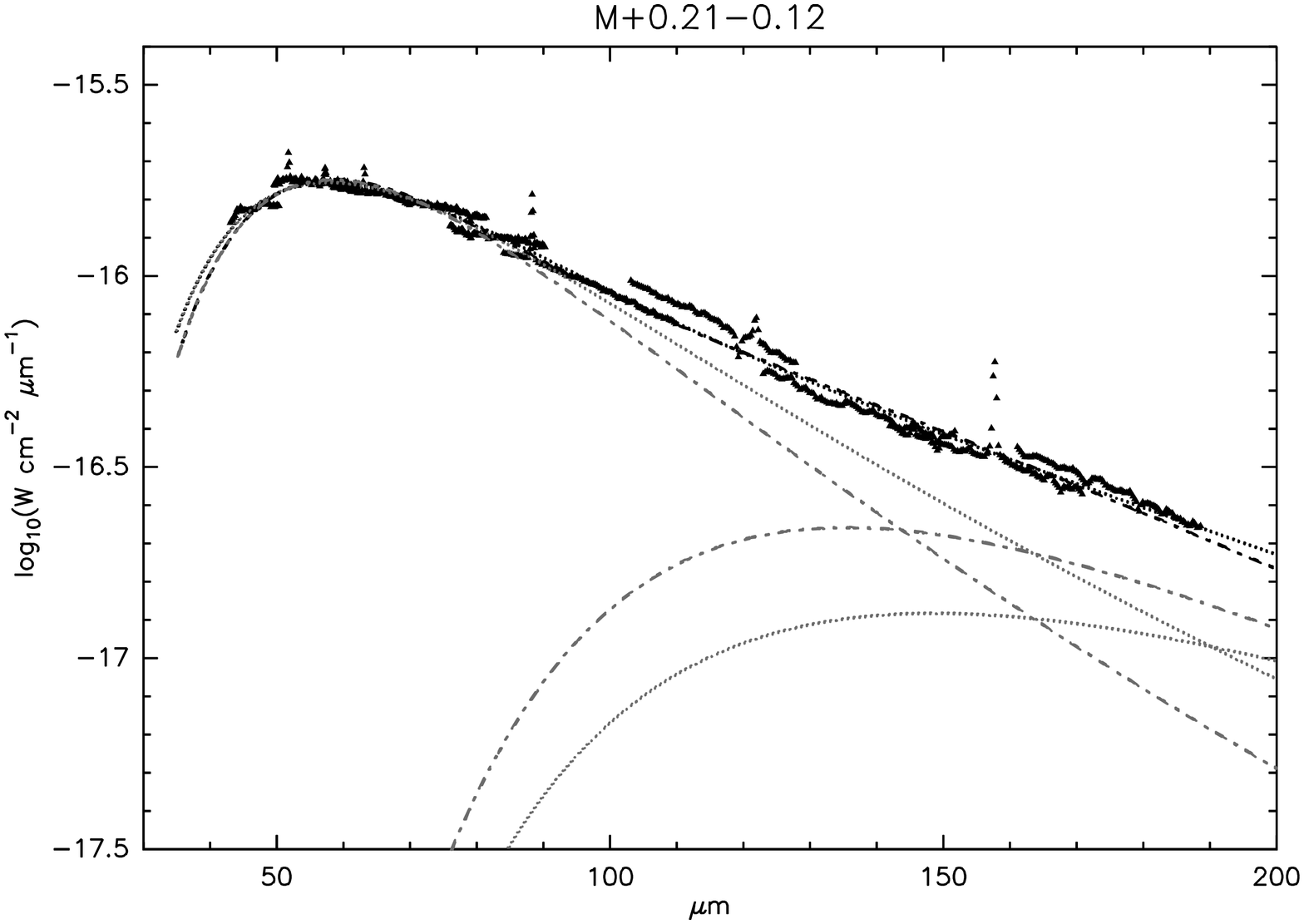}
\includegraphics[angle=-90,width=7cm]{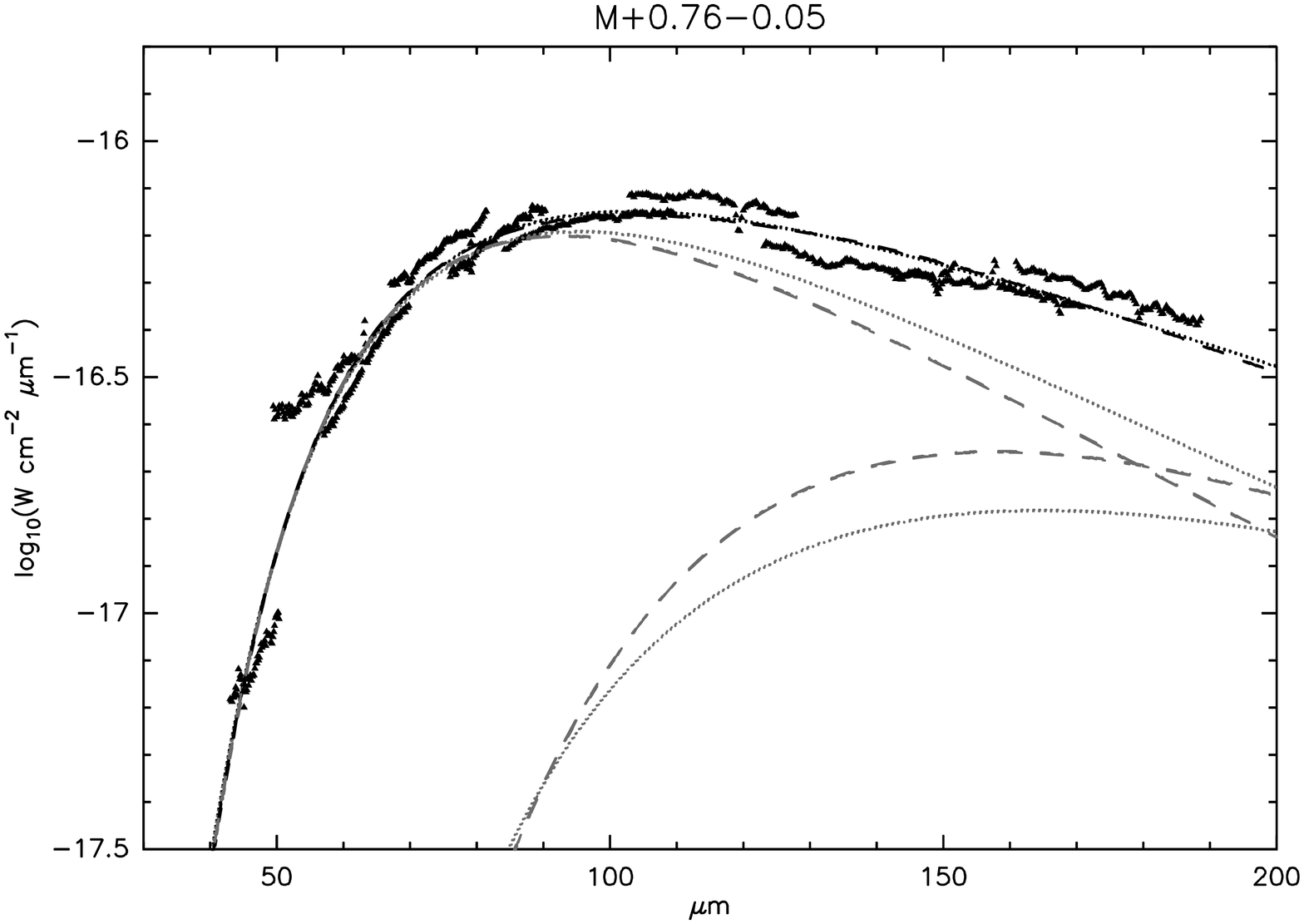}
\caption{Dust continuum emission toward M+0.21-0.12 and  M+0.76-0.05. The black solid points represent the observed spectra. For each source we show two different fits with two independent gray bodies. For each fit we show the total emission in black lines and the emission due to the cold and warm components in gray lines. For M+0.21-0.12 the dotted lines represent the fit obtained with
$\beta_w=1$ and $\beta_c=2$ and the dot-dashed  lines that obtained with $\beta_w=2$ and $\beta_c=3$. For M+0.76-0.05 the dotted lines represent the fit obtained with $\beta_w=1$ and $\beta_c=1$ and the dashed  lines that obtained
with $\beta_w=1.5$ and $\beta_c=2.5$. The other parameters of the fits can be found in Table \ref{tab_dust}
}
\label{fig_dust_fits}
\end{figure}

In principle $\beta$ varies from 1 for amorphous  to 2 for crystalline silicates. Using sub-millimeter data, Pierce-Price et al. (2000) and  Lis et al. (2001) have found higher $\beta$ values (2.5-3.4) in some GC sources. These authors interpret their finding as a consequence of the presence of icy mantles in the dust grains. We have fitted the dust temperatures ($T_w$ and $T_c$) and the opacities ($\tau_{30w}$ and $\tau_{30c}$) of the two dust components for different values of $\beta$ in the range of 1-3. Table \ref{tab_dust} and Fig. \ref{fig_dust_fits} show the results of fitting the spectrum towards M+0.21-0.12 (source with the hottest dust) and M+0.76-0.05 (source with the coldest dust). Fits of similar quality (as determined from $\chi^2$) can be obtained for $\beta$ values in the range of 1-3. The best fits have been obtained with $\beta=1$ for both the cold and the warm components. Fits with higher $\beta$'s present $\chi^2$ values just $\sim 15 \%$ higher. 

The fit results for $\tau_{30w}$ and $\tau_{30c}$ are strongly dependent on $\beta_c$ and  $\beta_w$. However, the total opacity is dominated by the cold dust component ($\tau_w/\tau_c \sim 1/50$; Table \ref{tab_dust}). On the other hand, one should note that fitting the data with high $\beta$s requires very high dust opacities ($\tau_{30} \sim 100$) that imply a visual extinction, $A_V$, of $\sim 1000$ ($\tau_{30} = 0.014 \, A_V$; Draine 1989). These huge extinctions are orders of magnitude higher than those estimated from any other tracer (Rodríguez-Fernández and Martín-Pintado 2004 have found values of $A_V$=30-60). Therefore, in the observed wavelength range one can rule out $\beta$ values equal or higher than 2. This result is in agreement with that obtained by Goicoechea et al. (2004) for the Sgr B2 envelope. The high $\beta$ values found with sub-millimeter observations may be due to an intrinsic dependence of $\beta$ with the wavelength (Bernard et al. 1999).

Table \ref{tab_dust} also shows that in contrast to $\tau_{30}$, $T_w$ and $T_c$ are almost independent of $\beta_c$ and $\beta_w$ and can be determined within 2 or 3 K. Therefore, Table \ref{tab_dust_fits} only contains the dust temperatures of the cold and warm components in all the sources for $\beta$s in the range of 1-3. The temperature of the cold dust component is 14-18 K for all the  sources. In contrast, the temperature of the warm component varies from 23-25 K for M+0.76-0.05 to 36-42 K for M+0.21-0.12. We conclude that the GC clouds exhibit two dust components, one with a temperature of 14-20 K and another warmer component whose temperature changes from 25 to 40 K source to source. The cold  temperature component corresponds to the well-known cold dust in the GC (Odenwald \& Fazio 1984, Cox \& Laureijs 1989, Pierce-Price et al. 2000). On the contrary, the detection of a new warm dust component with a temperature up to 40 K is a goal of the ISO/LWS spectra  (see also Lis \& Menten 1998, Rodriguez-Fernandez et al. 2000, Lis et al. 2001, Goicoechea et al. 2004). Table \ref{tab_dust_fits} also shows the far-infrared flux of each dust component. It is noteworthy that, contrary to the total opacity, the total dust luminosity is dominated  by the warm dust contribution (90~$\%$; Table \ref{tab_dust_fits}).

%................................................
\begin{table}[h]
\caption[]{Detailed results of the fits to the dust emission in two sources with
two independent gray bodies (see text)}
\label{tab_dust}
\begin{tabular}{lllllll}
\hline
\noalign{\smallskip}
$T_w$ & $\tau_{30\,w}$ & $\beta_w$ & $T_c$ & $\tau_{30\,c}$ & $\beta_c$ & $\chi^2$ \\
\noalign{\smallskip}
\hline
\hline
\multicolumn{7}{c}{M+0.76-0.05} \\
\hline
\noalign{\smallskip}
25   & 0.27 & 1    & 13.9 & 101  & 3   & 162  \\
25   & 0.27 & 1    & 14   & 16.8 & 2   & 158\\
25   & 0.27 & 1    & 15   & 1.83 & 1   & 157 \\
24   & 0.60 & 1.5  & 15   & 6.3  & 1.5 & 166 \\
24   & 0.63 & 1.5  & 14   & 59.6 & 2.5 & 168 \\
23   & 1.63 & 2    & 13.4 & 58.1 & 2.0 & 180 \\
23   & 1.70 & 2    & 13.2 &174.3 & 2.5 & 170 \\
\hline
\multicolumn{7}{c}{M+0.21-0.12} \\
\hline
41.8   &  0.035 & 1   & 16.8 & 0.70 & 1   & 37 \\
41.6   &  0.036 & 1   & 14.6 & 9.04 & 2   & 38 \\
41.3   &  0.038 & 1   & 13.4 & 91.2 & 3   & 40 \\
38.6   &  0.082 & 1.5 & 15.4 & 18.9 & 2.5 & 44 \\
39.0   &  0.076 & 1.5 & 17.5 & 1.68 & 1.5 & 40 \\
36.0   &  0.183 & 2   & 16.7 & 6.17 & 2   & 41 \\
35.6   &  0.203 & 2   & 15   & 63.6 & 3   & 47  \\
\noalign{\smallskip}
\hline
\end{tabular}
\end{table}
%..................................

%.......................

%................................................
\begin{table}[htb]
\caption[]{Total flux (FIR) and temperature (T) of the warm ($w$) and cold ($c$) components needed to fit the spectra with two independent gray bodies.}
\label{tab_dust_fits}
\begin{tabular}{lllll}
\hline
\noalign{\smallskip}
Fuente & FIR$_w$ & FIR$_c$ & T$_c$ & T$_w$  \\
       & W\,cm$^{-2}$ & W\,cm$^{-2}$ & K & K \\
       &$\times 10^{-15}$ & $\times 10^{-15}$ &    \\
\noalign{\smallskip}
\hline
\noalign{\smallskip}
M-0.96+0.13 & 2.4 & 0.5  & 14-15 & 28-31  \\
M-0.55-0.05 &10.6 & 1.7  & 15-16 & 30-33  \\
M-0.50-0.03 & 9.7 & 1.0  & 16-18 & 33-34  \\
M-0.42+0.01 & 8.6 & 1.2  & 16-17 & 29-32  \\
M-0.32-0.19 & 7.3 & 0.4  & 15-16 & 30-39  \\
M-0.15-0.07 & ... & ...  & ...   & ...    \\
M+0.16-0.10 & 7.8 & 2.1  & 16-17 & 30-33  \\
M+0.21-0.12 &12.1 & 1.4  & 16-17 & 36-42  \\
M+0.24+0.02 &12.9 & 3.2  & 16-17 & 29-31  \\
M+0.35-0.06 &13.4 & 1.1  & 16-20 & 33-39  \\
M+0.48+0.03 & 8.0 & 1.8  & 14-15 & 25-27  \\
M+0.58-0.13 & 6.6 & 0.6  & 14-16 & 26-28  \\
M+0.76-0.05 & 7.0 & 1.1  & 14-16 & 23-25  \\
M+0.83-0.10 & 5.5 & 0.8  & 15-16 & 25-28  \\
M+0.94-0.36 & ... & ...  &...    &  ...   \\
M+1.56-0.30 & 0.8 & 0.7  & 16-18 & 26-29  \\
M+2.99-0.06 & 0.7 & 1.0  & 16-19 & 27-31  \\
M+3.06+0.34 & 0.3 & 0.8  & 17-20 & 31-36  \\
\noalign{\smallskip}
\hline
\end{tabular}
\end{table}
%......................................

\begin{figure}[htb]
\centering
\includegraphics[angle=-90,width=6cm]{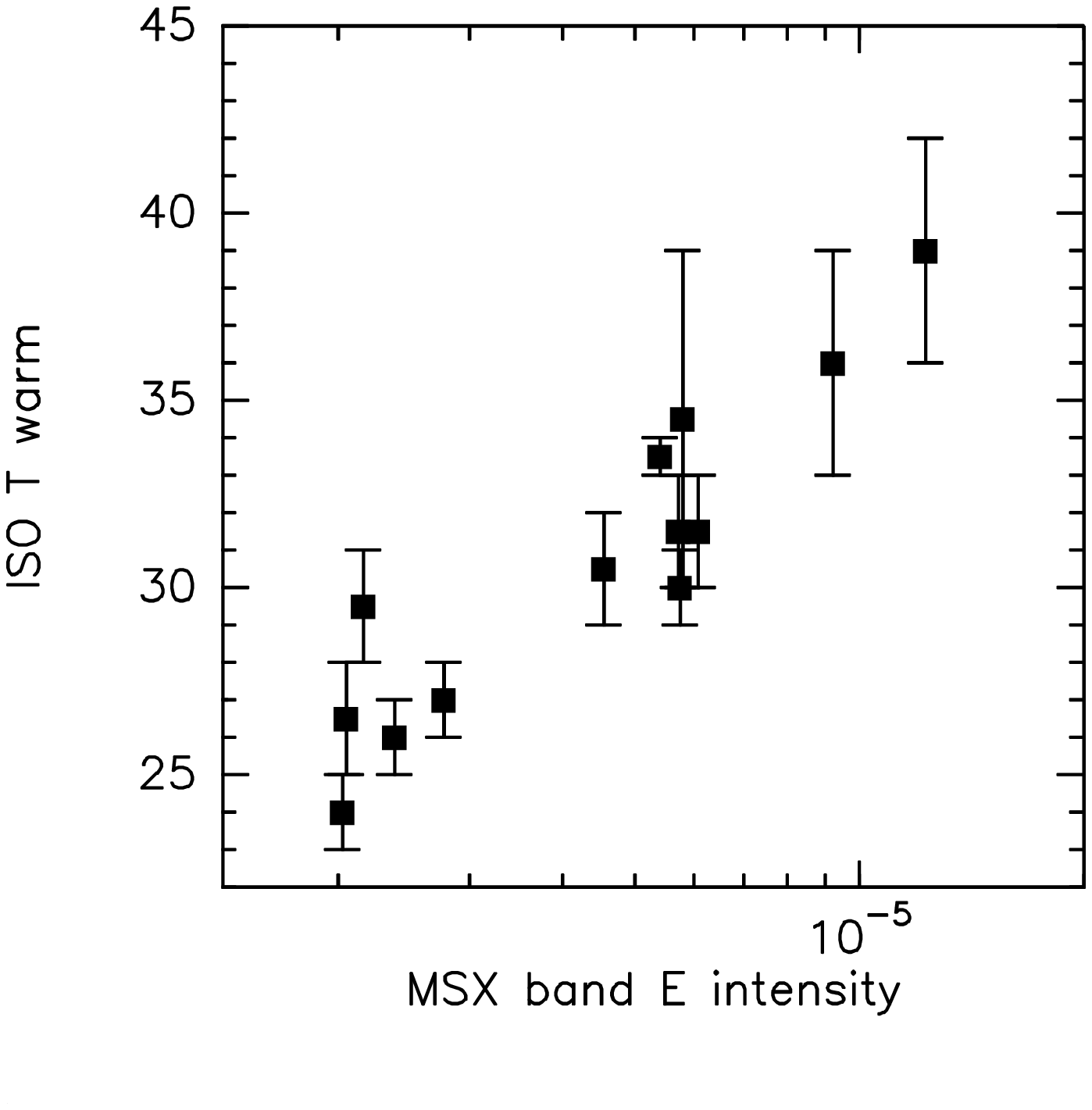}
\caption{Temperature (K) of the warm grey body needed to fit the dust spectra versus the MSX band E (18.2-25.1 \mum) intensity toward the sources presented in this paper}
\label{fig_t_msx}
\end{figure}

To get further insight on the distribution and properties of the warm dust in the GC we have compared our results with the   E band (18.2-25.1 \mum) image of the GC taken by the MSX satellite (see Fig. 1). This image shows extended emission over the central 300 pc of the Galaxy in addition to a number of point sources. We have averaged the MSX E band  intensity at the location of our sources in an area comparable to the LWS beam. In Fig. \ref{fig_t_msx}, we have plotted the temperature of the warm dust component derived from our fits versus the MSX E band intensity. The plot shows a good correlation between both quantities, which imply that the MSX image is showing the distribution of dust with a temperature higher than 30 K.  
Apart from the well-known radio continuum sources (Sgr A, B, C...), the warmest dust is found in the Thermal Filaments and the Sickle nebula (both in the Radio Arc region), in a source located at $(l,b)\sim(0.3,-0.01)$ and towards $(l,b)\sim(-0.35,-0.03)$.

%.....................................................................................
\section{Fine structure lines}

\subsection{Comparison with other observations}

\begin{figure*}[htb]
\includegraphics[angle=-90,width=14cm]{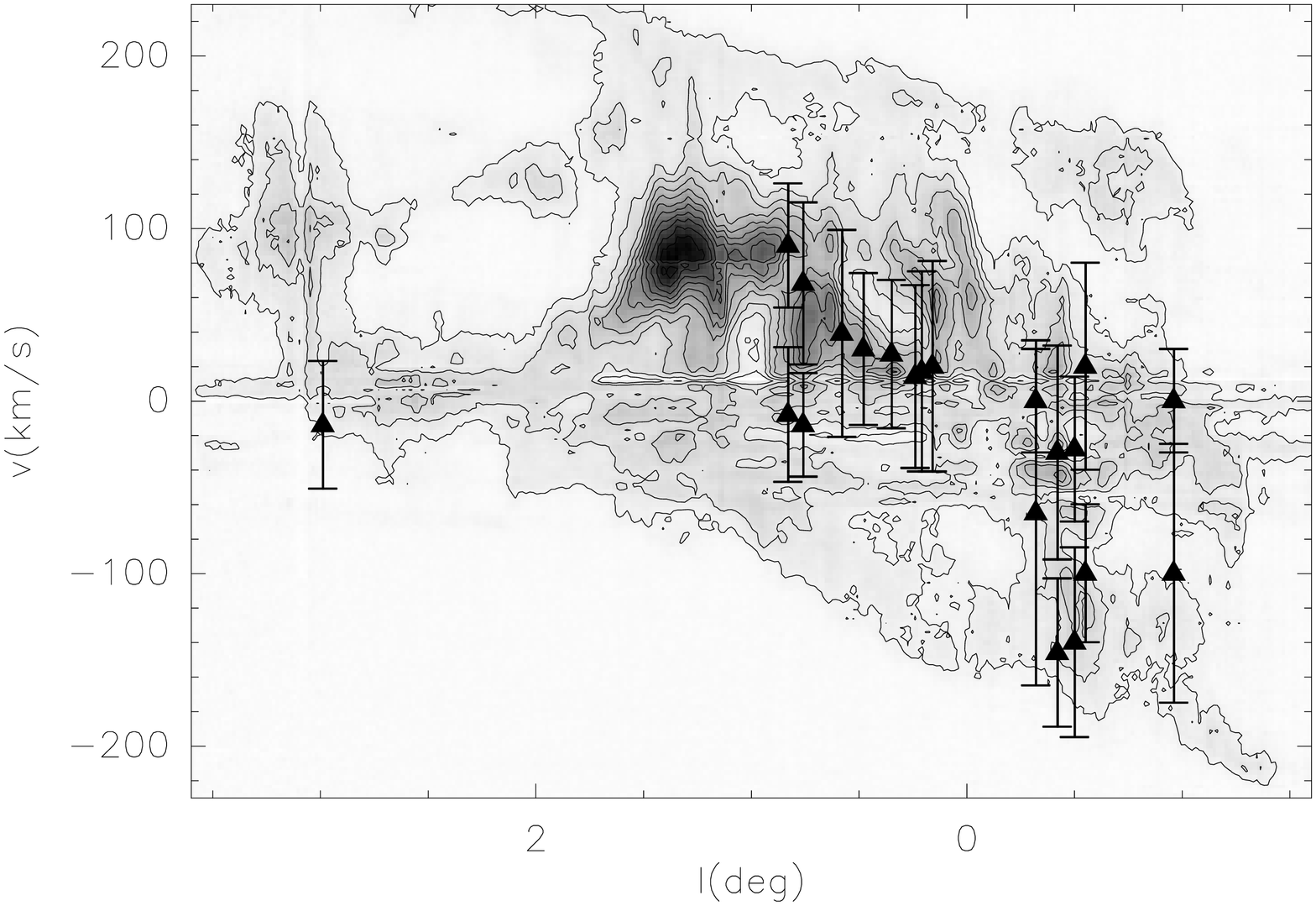}
\caption{Peak velocities (triangles) and linewidths (errorbars) of the [\CII] 158 \mum~ line overplotted on the CO(1-0) longitude-velocity diagram by Bally et al. (1987)}
\label{fig_CII_lv}
\end{figure*}

The high spectral resolution of the Fabry-Perot observations allows us to study the line profiles and peak velocities of the fine-structure lines, especially  those of the [\CII] lines, which have a high signal-to-noise ratio. To compare the [\CII] data with molecular lines observations, in Fig. \ref{FPvsCO_all.eps} we overplot the [\CII] lines on the $^{13}$CO(1-0) spectra of Rodríguez-Fernández et al. (2001a). In addition, in Fig. \ref{fig_CII_lv} we show the central velocities  of the [\CII] lines (solid triangles) and the linewidths (errorbars) derived from the Gaussian fits (Table \ref{tab_lws04}) overlaid in the $^{12}$CO(1-0) map of Bally et al. (1987).

First, it is noteworthy that the [\CII] emission in M-0.96+0.13 arises in two components with velocities of 0 and -100 \kms ~ but no emission is detected from the prominent CO component at 150 \kms. In the low spectral resolution H$_2$ spectra of Rodríguez-Fernández et al. (2001a) we also found evidences that the warm H$_2$ emission arises mainly from the negative velocity gas. Thus, the warm H$_2$ and the [\CII] emission are probably associated.  The sources in the Sgr C area (M-0.42+0.01, M-0.50-0.03 and M-0.55-0.05)  show clearly two [\CII] velocity components. This is in agreement with the hydrogen recombination  lines observations of Pauls \& Mezger (1975). The [\NII] and [\OIII] data of Rodríguez-Fernandez and Martín-Pintado (2004) also show both velocity components. However, in contrast to the [\CII], where both components have  a similar intensity, the [\NII] and [\OIII] lines are much more prominent in the v=-100 \kms ~ component, which is gas associated to the Sgr C \HII ~ regions.

In the Radio Arc region (M+0.21-0.12 and M+0.16-0.10)  the [\CII] lines are centered at $\sim 10$ \kms. Thus,  these seem to arise in the same gas that was detected by Pauls \& Mezger (1980) in hydrogen recombination lines. That velocity is also in agreement with the [\OIII] and [\NII] data of Rodríguez-Fernandez and Martín-Pintado (2004). On the contrary, the CO emission arises mainly at 60 \kms ~ (Figs. \ref{FPvsCO_all.eps} and \ref{fig_CII_lv}). Therefore, the bulk of the [\CII] emission in the Radio Arc is not associated to the dense molecular clouds. Unfortunately, the low spectral resolution H$_2$ data of Rodríguez-Fernández et al. (2001a)  do not allow us to discriminate the origin of the warm H$_2$ emission. Nevertheless, the warm NH$_3$ emission observed by Hüttemeister et al. (1993) in the only source of their sample located in the Radio Arc area (M+0.10-0.12) arises in gas with a velocity of 26 \kms, between that of the ionized gas and the 60 \kms ~ CO cloud.

For the rest of the sources the agreement between peak velocities and line profiles of the [\CII] and CO lines is quite good (see for instance M+0.83-0.10 or M+2.99-0.06). However, for some sources the fine structure lines peak velocities show  a small shift respect to the CO lines (see for instance M+0.48+0.03 and M+0.76-0.05 in Fig.  \ref{FPvsCO_all.eps}).  The magnitude of this shift is similar to the Fabry-Perot spectral resolution (30 \kms) but larger than the wavelength calibration uncertainties (7 \kms), therefore we believe it is real. The velocity shift is also obvious in Fig. \ref{fig_CII_lv} where the [\CII] peak velocities are always in the envelopes of the dense molecular clouds.  This finding suggests that the clouds are illuminated from the outside and that the [\CII] emission arises in PDRs in the external layers of the clouds. This is in agreement with the recent results from the AST/RO survey by Martin et al. (2004). Their longitude-velocity diagrams show that the kinetic temperatures of the cloud envelopes are higher than  those of the central parts of the clouds.

\subsection{Extinction}

The fine structure lines presented in this paper have been corrected for extinction following the method  discussed in detail in Rodríguez-Fernández and Martín-Pintado (2004). Basically, a lower and an upper limit to the extinction have been derived for every source. The lower limits were obtained from the [\SIII] 18/33 \mum ~ lines ratios and are similar to the foreground extinction to the Galactic center as derived from star counts (A$_V \sim 25$ mag). On the other hand, conservative upper limits to the extinction have been obtained by comparing radio and infrared hydrogen recombination lines as well as from the total column density of molecular gas measured toward each source. These limits range from
$\sim 45$ to $\sim 90$ mag ($A_V$) from source to source (see Rodríguez-Fernández and Martín-Pintado 2004). All the following analysis has been done taking this range of possible extinctions into account. The extinction correction uncertainties are the main source of errors in the results. To obtain the actual values of the extinction at the  wavelength of a given line, we have used the Draine (1989) law in the mid infrared range. For the lines in the far-infrared range, we used the 30 \mum ~ extinction from Draine (1989). We have extrapolated to longer wavelengths using a power law with exponent 1.5.

\subsection{Excitation: shocks versus radiation}

\begin{figure*}[tbh]
\includegraphics[width=17cm]{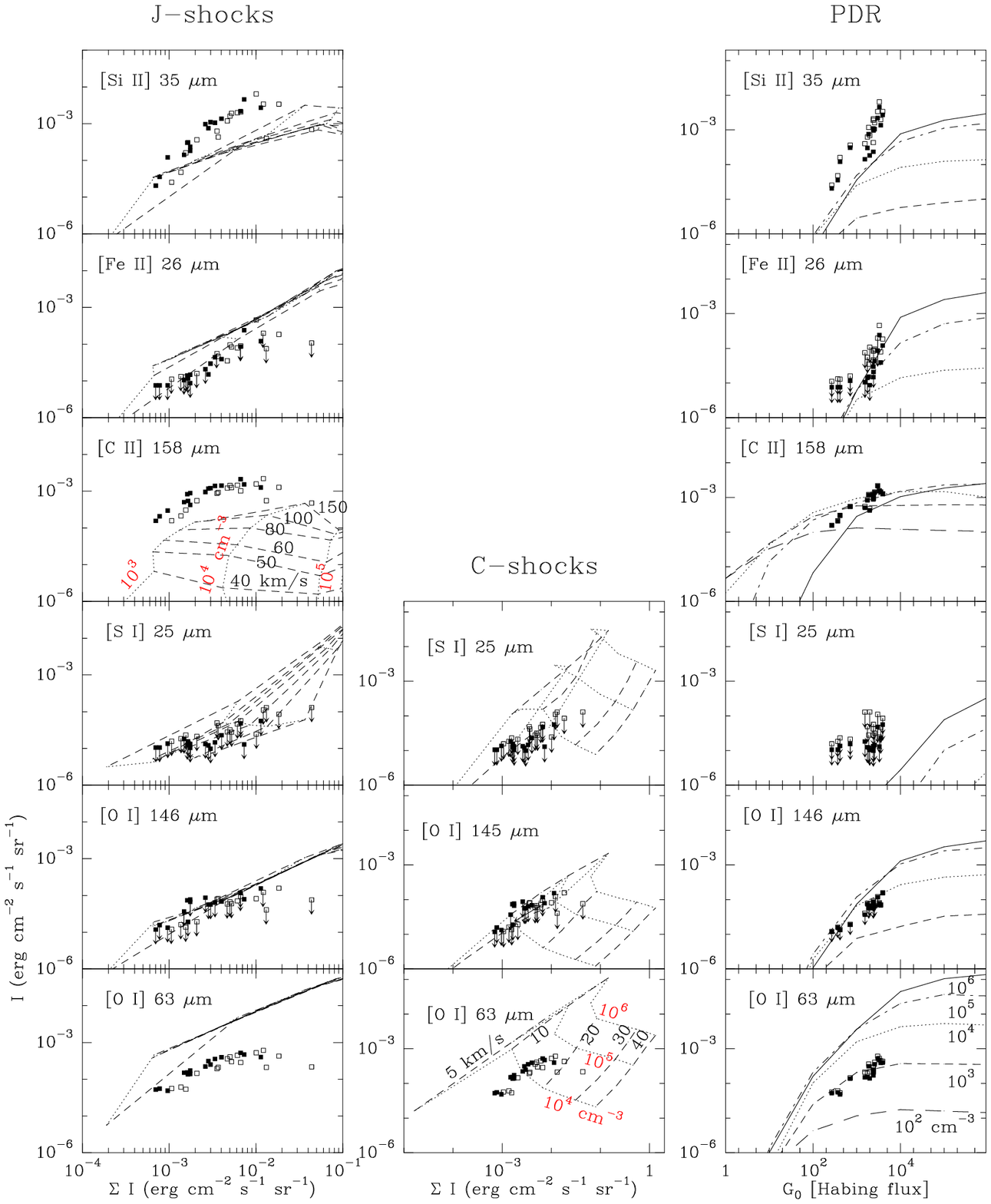}
\caption{ {\bf Left panels:} Fine-structure lines intensities versus the total intensity radiated by lines. Solid and empty squares represent the observed results for the low and high extinction corrections, respectively. For comparison, the lines intensities versus the total intensity radiated by lines in J-shocks of different velocities (dashed lines) and densities (dotted lines) are also shown.  {\bf Central panels:} Same for C-shocks of different velocities (dashed lines) and densities (dotted lines). {\bf Right panels: }  Intensities of different lines versus the far-ultraviolet incident flux. Solid and empty squares represent the results for the low and high extinction corrections, respectively. For comparison, the PDR model predictions for different densities are also shown}
\label{fig_fe}
\end{figure*}

The lines presented in this paper are among the main coolants of the interstellar gas with a temperature of a few hundred K. Thus, the study of this line emission (in addition to the dust continuum of the previous sections) can give us important estimates of  the thermal balance of the GC clouds and the heating mechanisms of the warm gas. In the following, we perform a similar analysis to that done by Van den Ancker (1999) to study the ISO spectroscopy of Young Stellar Objects (Fig \ref{fig_fe} has been made with data kindly provided by Mario Van den Ancker).

\subsubsection{Shocks}
The left panels of Fig. \ref{fig_fe} shows the J-shock model predictions (Hollenbach \& McKee 1989) for the intensity of several lines as a function of the total intensity emitted by all lines for different shock velocities (30-150 \kms) and preshock gas densities (10$^3$-10$^6$ \cmmt). J-shocks models fail to predict the observed line intensities: first, the observed [\OI] 63 \mum ~ intensity is considerably lower than that predicted by the models, and second,  J-shocks models do not predict enough emission from  [\CII] 158 \mum ~ and [\SiII] 35 \mum ~ even for shock velocities as high as 150 \kms.

The center panels of Fig. \ref{fig_fe} present C-shock model predictions (Draine et al. 1983) for the intensity of several lines as a function of the total intensity emitted by all lines for shocks with velocities of 5-40 \kms ~ and preshock gas densities of 10$^4$-10$^6$ \cmmt. The [\OI] 63 \mum ~ intensities can be explained by a C-shock with a preshock density in the range of 10$^3$-10$^5$ \cmmt ~ and shock velocities of 10-20 \kms. The detected  [\OI] 146 \mum~ intensities as well as the upper limits suggest densities somewhat higher (10$^4-10^5$ \cmmt) than those estimated with the [\OI] 63 \mum~ line and slightly lower velocities (5-10 \kms). The [\SI] lines seem to rule out C-shocks in gas with a pre-shock densitiy higher than 10$^5$ \cmmt. In summary, fine structure lines from neutral species can arise in C-shocks with 10 \kms ~and 10$^{4}$ \cmmt.

\subsubsection{Diffuse ionized gas and PDRs}
The right panels of Fig. \ref{fig_fe} show the intensities of several lines versus the incident far-ultraviolet flux as predicted  by PDR models (Tielens \& Hollenbach 1985) with gas densities in the range of 10$^2$-10$^6$ \cmmt. As discussed by Van den Ancker (1999), in PDRs the far-infrared intensity ($I_{FIR}$) is correlated to the far-ultraviolet incident intensity  ($I_{FUV}$) by the relation

\begin{equation}
I_{FUV}=0.5\frac{I_{FIR}}{\Omega^2}
\end{equation}

where $\Omega$ is the spatial extent of the PDR. Here we have assumed that the PDRs fill the LWS beam. To obtain the intensity in terms of $G_0$, we have divided $I_{FUV}$ by the average local interstellar FUV field (1.2\,10$^{-4}$ erg\,\smu\,\cmmd\,sr$^{-1}$, Habing 1968).

The [\FeII] 26 \mum ~ and [\SI] 25 \mum ~ intensities and upper limits are compatible with the predictions of the PDR models. The intensity of the [\OI] lines can be explained with a density of 10$^3$-10$^4$ \cmmt ~ and a FUV incident field $G_0$  of  ~ $\sim 10^{3.3}$ for all the sources except those located in the extremes of the Central Molecular Zone (CMZ, Morris \& Serabyn 1996) and those located in the Clump 2 (M-0.96+0.13, M+1.56-0.30, M+3.06+0.34 and M+2.99-0.06). The line emission in these  sources would arise in PDRs of a similar density (10$^3$ \cmmt) but with a lower $G_0$ ($\sim 10^{2.5}$). That is, the interstellar radiation field would be around 10 times lower than in the internal regions of the CMZ (from Sgr C to Sgr B2). The intensity of the [\CII] 158 \mum ~  line can also be naturally explained in the context of a 10$^2$-10$^5$ \cmmt ~ PDR. However, the PDR model prediction for the [\SiII] 35 \mum ~ line intensity is $\sim 10$ times lower than observed.

\begin{figure}[tbh]
\includegraphics[angle=270,width=8cm]{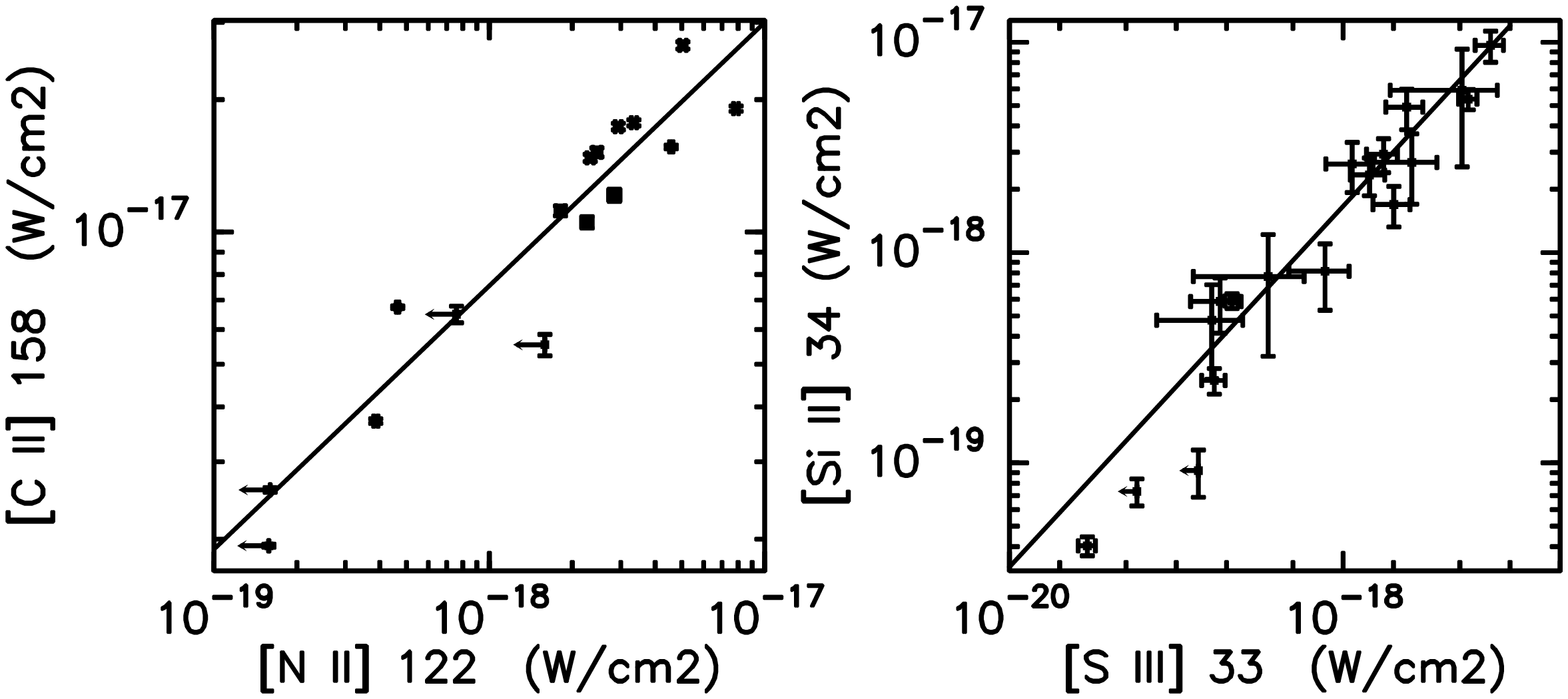}
\caption{Left: [\CII] 158 \mum ~ versus [\NII] 122 \mum ~ line
fluxes. Right: [\SiII] 34 \mum ~ versus [\SIII] 33 \mum ~ line
fluxes }
\label{fig_correl}
\end{figure}

The ionization potential of C{\sc i} and Si{\sc i} is lower than 13.6 eV. Therefore, [\CII] and [\SiII] lines can arise from mainly neutral as well as from ionized regions. The photo-ionized gas component presented in Rodríguez-Fernández and Martín-Pintado (2004) can also contribute to the flux of these lines. To investigate this possibility in Fig. \ref{fig_correl} we have plotted  the [\CII] line flux versus the [\NII] flux (both lines have been observed with the LWS beam) and the [\SiII] line versus the [\SIII] 33 \mum ~ line (both observed with the same SWS beam). The two plots show remarkable correlations. Contrary to the [\SiII] and [\CII] lines, the [\NII] and [\SIII] can only arise in the ionized gas (the ionization potential of  N and S$^{+}$ is higher than 13.6 eV). Therefore, the observed correlations imply that a fraction of the [\CII] and [\SiII] fluxes should indeed come from the ionized gas.

A least squares fit to the [\CII] and [\NII] data gives the following result:

\begin{equation}
F(CII)=(3.8\pm0.6)F(NII)+(4\pm2)10^{-18}
\end{equation}

where $F(CII)$ and $F(NII)$ are the [\CII] 158 and [\NII] 122 \mum ~ fluxes in W\,\cmmd. For comparison, Goicoecha et al. (2004) found an slope of 5.2 in the Sgr B2 envelope and Malhotra et al. (2001) found an slope of 4.3 in their sample of normal galaxies. Since the bulk of the [\NII] emission arises in the extended low density ionized gas, we conclude that a fraction of the [\CII] emission also arises in this gas component. One can approximate the [\CII] flux arising from the PDR, $F(CII_{PDR})$, as $F(CII_{PDR})=F(CII)-3.8\,F(NII)$ which is on average of $4\,10^{-18}$ W\,\cmmd. Therefore, for the Clump 2 and the outer CMZ sources the [\CII] flux is likely to be mostly of PDR origin while for the inner CMZ sources up to $\sim 70$ \% of the flux could arise in  extended low density ionized gas.

On the other hand, a least squares fit to the [\SiII] and [\SIII] fluxes gives

\begin{equation}
F(SiII)=(1.11\pm0.0.09)F(SIII)+(4\pm3)10^{-19}
\end{equation}

where $F(SiII)$ and $F(SIII)$ are the [\SiII] 35 \mum~ and [\SIII] 33 \mum ~ fluxes in W\,\cmmd. To understand this relation we have computed some CLOUDY (Ferland 1996) models. We have studied the  [\SiII]/[\SIII] ratio in a range of electron densities (0.1-10$^4$ \cmmt), effective temperatures of the ionizing radiation (25000-40000 K) and ionization parameters ($logU=-1,-4$). In that range of parameters,  the [\SiII]/[\SIII] ratio is almost independent of the electron density and only weakly dependent on the radiation temperature. The predicted [\SiII]/[\SIII] ratio in the ionized gas is $\sim 0.2-1$ for ionization parameters between $logU=-2$ and $-3$, respectively, which is the range derived from the data of Rodríguez-Fernández and Martín-Pintado (2004). Since the observed correlation gives a [\SiII]/[\SIII] ratio in the range 1.0-1.2 one concludes that most of the [\SiII] emission could indeed arise in the ionized gas. This result explains naturally the observed [\SiII] excess when compared to the PDR models. Alternatively, the silicon gas phase abundance in the GC could be much higher than that considered by the PDR models ($\sim 3.6\,10^{-6}$). However, to explain the observed flux excess all the silicon should be in gas phase. We conclude that a fraction of the [\SiII] flux comes from the ionized gas. For NGC 7023, this result is in agreement with the fact that the [\SiII] line is more intense toward the ionized gas than toward the PDR itself (Fuente et al. 2000).

In summary, the fine structure lines of neutral species are consistent with a C-shock or a PDR. However, the lines from ionic species imply a PDR/ionized gas  scenario.

%..................................................................
\section{H$_2$ pure rotational lines}

The H$_2$ pure rotational lines toward the GC clouds have been discussed by Rodríguez-Fernández et al. (2000, 2001a). The S(0) and S(1) lines trace a gas temperature of $\sim 150-200$ K while the lines up to the S(5) probe gas temperatures up to $\sim 500$ K. The column density of H$_2$ at $\sim$ 150 K is a few 10$^{22}$ \cmmd ~ while the column density of gas at 500 K is only $\sim 1\%$ of that at 150 K. On average, the column density of gas with a temperature of 150 K is $\sim 30 \%$ of the total gas column density estimated from CO but there are significant differences from source to source (the warm gas fraction in M-0.96+0.13 is close to 1). The H$_2$ excitation was also discussed by Rodríguez-Fernández et al. (2001a) by comparing the population diagrams derived from the observations with the same type of diagrams constructed from J- and C-shock and PDR models. The results were that the emission from lines arising from more excited levels than those connected to the S(1) line are consistent with   a C-shock of 12 \kms ~ and 10$^6$ \cmmt, a PDR of $G_0=10^4$ and  10$^6$ \cmmt ~ (both comparing with models and observational templates as NGC7023) or a J-shock of 50 \kms~ and 10$^6$ \cmmt. In contrast, to explain the S(0) and S(1) intensities several PDRs with lower density and incident field ($G_0=10^3$, $n=10^3$\cmmt) or several C-shocks with velocities as low as 7 \kms ~ are required. Of course, a combination of these PDRs and low velocity shocks is  also  possible.

\begin{figure*}[thb]
\includegraphics[width=18cm]{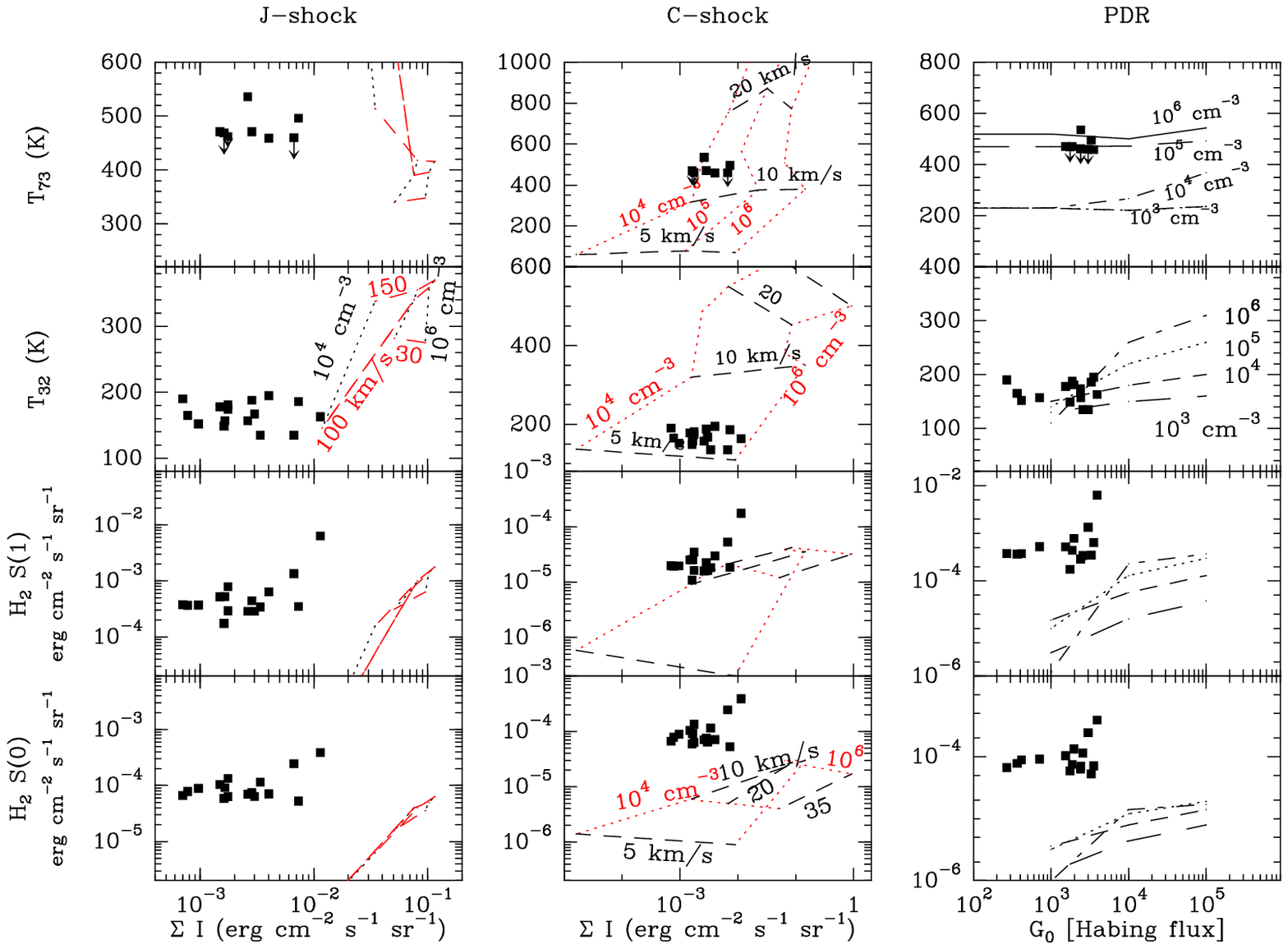}
\caption{Left panel: H$_2$S(0) and H$_2$S(1) intensities and rotational temperatures derived from the J=3 and J=2 levels ($T_{32}$) and from the J=3 and J=7 levels ($T_{73}$). All those parameters are plotted versus the total intensity radiated by lines. For comparison, the expected results for J-shocks models with different densities (dotted lines) and velocities (dashed lines) are also shown.  Middle panel: Same for C-shocks. Right panel: H$_2$S(0), H$_2$S(1), $T_ {32}$ and $T_{73}$ versus $G_0$ and comparison with PDR models with different densities}
\label{fig_h2}
\end{figure*}

In this section we revisit the H$_2$ excitation in the same context in which we
have studied the fine structure line emission in the previous section. Figure \ref{fig_h2} shows the intensity of the S(0) and S(1) lines as well as the excitation temperatures derived from the J=3 and J=2 levels ($T_{32}$) and the J=3 and J=7 levels ($T_{73}$). These quantities are represented versus the total intensity radiated by lines and compared to different  J-shocks (left panel) and C-shocks (middle panel) model predictions. The right panel of Fig. \ref{fig_h2} presents our S(0), S(1), $T_{32}$ and $T_{73}$ values versus $G_0$ (derived from the far-infrared flux). We also show a comparison with PDR model predictions.

The H$_2$ emission is difficult to study in J-shocks due to the dissociation and subsequent molecule reformation. Furthermore there are no model predictions for the low-J H$_2$ pure rotational lines in shocks with preshock density lower than 10$^4$ \cmmt. The observed  $T_{73}$ and $T_{32}$ are in the range of possible values expected by J-shock excitation. However,  the total intensity radiated by lines in a J-shock with preshock density higher than 10$^4$ \cmmt ~ is much higher than that observed from the GC sources. In addition, the measured line intensities of the S(0) and S(1) lines would be at least 2 orders of magnitude higher than those predicted by these models.  In summary, J-shocks models cannot explain the observed H$_2$ emission and the total energy radiated by lines.

The central panels of Fig. \ref{fig_h2} show that both $T_{73}$ and $T_{32}$ can be explained by C-shocks models with densities of 10$^4-10^6$ \cmmt ~ and velocities of 5-15 \kms. Moreover, the total intensity radiated by lines in such shocks is in agreement with the observations. Figure \ref{fig_h2} also shows that the observed S(1) intensities are around 3 times higher than  those predicted by a 10$^4$ \cmmt ~ and $v=10$ \kms ~ shock. This effect is much stronger for the S(0) line since the observed intensities are around a  factor of 10 higher than those predicted by the models. 

The right panels of Fig. \ref{fig_h2} show that both  $T_{73}$ and $T_{32}$ can be explained in PDRs of $n=10^3-10^5$ \cmmt. Moreover the observed far-infrared flux due to the warm dust (and $G_0$) is also consistent with this picture. However, the observed S(1) line intensity is  a factor of $\sim 10$ higher than  predicted by a PDR model with $G_0=10^3$ and $n=10^6$ \cmmt. The discrepancy between  observations and models is even higher for the S(0) line. 

In summary,  the H$_2$ emission from excited levels can be explained either by a C-shock or a PDR. However, neither C-shocks nor PDR models can account for the large observed intensities of the S(0) and S(1) lines.

\section{Discussion: the origin of the heating and the chemistry of the GC clouds}

\subsection{Photo-dissociation regions}

In the previous sections we have shown that the GC clouds exhibit a warm dust component with a temperature up to 45 K that sould be heated by UV radiation. Regarding the gas, on the one hand, the fine structure lines of neutral species and the  H$_2$ emission from excited levels could be heated by a C-shock or a PDR. On the other hand, the fine structure lines from ionized species are best explained in a PDR/DIG (diffuse ionized gas) scenario. The large scale observations of the H$_2$  $v=1-0$ S(1) ro-vibrational line at 2.12 \mum ~ by Pak et al. (1996) has also been interpreted as arising in a $G_0=10^3$ and $n=10^3$\,\cmmt ~ PDR. These parameters are in excellent agreement with those derived in the previous sections. Clearly, the extended PDR/DIG scenario is the easiest way to explain all the observational data. Namely, the warm dust, the fine structure lines from neutral and ionized gas, and the H$_2$ emission from excited levels. 

As discussed in Rodríguez-Fernández and Martín-Pintado (2004), the interstellar medium in the 300 central pc of the Galaxy is quite non-homogeneous and it is permeated by relatively hot stellar radiation ($\sim 35000$ K). The DIG emission seems to arise from the external layers of the dense molecular clouds. Indeed, molecular, neutral/atomic and ionized gas are associated (Sect. 4.1) and therefore a PDR is expected in the interface between the dense molecular gas and the ionized gas. In the external layers of such a PDR, the gas is heated by photoelectric effect on the dust grains to temperatures well upon  the dust temperature. For instance the standard model of a $G_0=10^3$ and $n=10^3$\,\cmmt ~ PDR by Hollenbach et al. (1991) shows that in the first two or three magnitudes of visual extinction inside the PDR, the gas is heated to a temperature of 100-200 K while the dust temperature is 30-50 K. Thus, the discrepancy of gas and dust temperatures is normal in PDRs. Furthermore, the measured gas and dust temperatures in the GC clouds are in excellent agreement with the Hollenbach et al. (1991) models. However, the column density of molecular gas with a temperature of $\sim 150$ K predicted by these models is 2-3\,10$^{21}$ \cmmd, i.e., 10-30 $\%$ of the measured warm gas column density in the GC clouds (Rodríguez-Fernández et al. 2001a).
For some lines of sight the discrepancy could be lower if one takes into account that several clouds are observed (different velocity components) and each of them can present a PDR.
Nevertheless, even with 2 or 3 standard PDRs in the line of sight, it will be difficult to explain the large column densities of warm gas measured in  the GC clouds.

 In addition, Rodríguez-Fernández et al. (2001a) have measured a warm NH$_3$ abundance in the GC clouds of $\sim 10^{-7}$, which is only 4 times lower than the typical C$^{18}$O abundance in the GC (Wilson \& Matteucci 1992). This large NH$_3$ abundance is difficult to explain  in the context of a PDR. For instance, the NH$_3$ to C$^{18}$O abundance ratio measured in the Orion bar PDR is $\sim 10^{-2}$ (Batrla \& Wilson 2003), well below the value measured in the GC. In a PDR the warm NH$_3$ emission arises from the external layers, where the gas is more exposed to the radiation. The problem is that the NH$_3$ molecule is easily photo-dissociated by UV radiation. The NH$_3$ dissociation rate can be expressed as
$\sim 3\,10^{-10} G_0 \exp[-2.16\,A_V+1.71\,10^{-2} \,A_V^2]$ \smu ~
 (Stenberg \& Dalgarno 1995). Therefore the dissociation rate in the warm region ($A_V<2$ mag.) of a $G_0=10^3$ PDR  is $\sim 4.3\,10^{-9}$ \smu. Thus, there would be no warm NH$_3$ after $\sim 7$ yr. One can estimate whether the NH$_3$ abundances could be explained in a dynamical PDR in which NH$_3$ rich gas is continuously been injected in the photo-dissociation front. To maintain a constant NH$_3$ abundance of 10$^{-7}$, a NH$_3$ column density of 2\,10$^{14}$ \cmmd~ (H$_2$ column density of 2\,10$^{21}$ \cmmd) should be injected to the PDR every 7~yr. Taking into account a H$_2$ density of 10$^3$ \cmmt ~ one derives that the NH$_3$ rich gas should be entering the photo-dissociation front with a velocity of several thousand \kms, which is not observed. 

The large warm H$_2$ column density and the NH$_3$ abundance measured in GC clouds is difficult to explain in a PDR scenario, even if the photo-dissociation front is not stationary. In the following  we review other alternatives.

\subsection{Cosmic rays}

A  cosmic rays ionization rate of $\sim 10^{-17}$ \smu ~ maintains the gas temperature at $\sim 10$ K in the local quiescent molecular clouds. An ionization rate  100 times higher than the local value would balance the gas cooling rate at 70 K. Indeed, an  enhanced cosmic rays ionization rate was one of the  first mechanisms  proposed to explain the warm temperatures of the GC clouds (Güsten et al. 1981, Morris et al. 1983). However, it is now clear that the GC clouds present a wide rage of temperatures from 20-30 K to 150 and 500 K (Hüttemeister et al. 1993,   Rodríguez-Fernández et al. 2001a). Temperatures of $\sim 70$ K, which are sometimes cited as typical of the GC clouds, are just averaged values without much physical meaning.
Heating by cosmic rays is a very pervasive mechanism since cosmic rays penetrate and traverse the clouds. If the cosmic rays ionization rate is high enough to heat the gas to 150 K most of the gas in the GC clouds should present this temperature. On the contrary,  Rodríguez-Fernández et al. (2001a) showed that only 30$\%$ of the gas is warm. Furthermore,  EGRET observations do not show any excess of $\gamma$-rays associated to the GC molecular complexes  that could be due to a cosmic rate ionization rate higher than the local value (Mayer-Hasselwander et al. 1998). We conclude that cosmic rays cannot be the heating mechanism of the bulk of the 150 K gas.

\subsection{X-rays dominated regions (XDRs)}
The GC is an extended source of Fe 6.4 keV emission (Koyama et al. 1996). This line arises from neutral or weakly ionized gas illuminated by a hard X-ray continuum. Martin-Pintado et al. (2000) have found a spatial correlation of the SiO(1-0) and the Fe 6.4 keV emission that suggest that both lines have a related origin and therefore, it extends the possible influence of the X-rays to the molecular gas. The GC clouds could be X-ray Dominated Regions (XDRs) as those modeled by Maloney et al. (1996). X-rays act directly by photo-electric effect on the gas and their attenuation is directly proportional to the gas column. As a result, the column density of warm gas in XDRs can be 10 times higher than that of a  PDR (in PDRs the dominant gas  heating mechanism is the photo-electric effect on the dust grains and the UV absorption by the dust is exponential). Therefore, XDRs could heat large quantities of gas, such as those measured in the GC.

XDRs can be characterized by the effective ionization parameter ($\epsilon_{eff}$) and the gas density (Maloney et al. 1996). In clouds with a density of 10$^3$ \cmmt ~ and a column density of 10$^{22}$ \cmmd, $\epsilon_{eff}$ can be expressed as: $\epsilon_{eff}=1.1\,10^6 \, L_{34} \,r^{-2}$, where $L_{34}$ is the X-rays luminosity in units of 10$^{34}$ erg \smu ~ and $r$ is the distance from the cloud to the continuum source in pc. The X-rays luminosity in scales of 8$^{'}$ ($\sim 20$ pc) around SgrB2, C or D is $\sim 0.3\,10^{35}$ erg \smu ~ while around Sgr A it is approximately 3 times higher (Sidoli et al. 1999). Taking $L_{34}=100$ and a spatial scale of 10 pc one gets $\epsilon_{eff}=1.1\,10^{-6}$. In contrast,  $\epsilon_{eff} \gsim 10^{-3}$ is required to heat the gas to temperatures higher than 100 K (Maloney et al. 1996). Therefore, XDRs do not seem to be a possible heating mechanisms in large scales for the bulk of the gas.

\subsection{Turbulence or low velocity shocks}
The sound velocity in a gas with a temperature of $\sim 100$ K is $\sim 1$ \kms. Hence, the  large widths of the molecular lines observed towards the GC (10-15 \kms) imply the presence of a high degree of supersonic turbulence. In addition to the large line-widths and the widespread high gas temperatures, the GC clouds exhibit a particular hot-core-like chemistry. High abundances and extended emission of molecules like C$_2$H$_5$OH, SiO (Martín-Pintado et al. 1997, 2000, 2001), NH$_3$ (Hüttemeister et al. 1993,   Rodríguez-Fernández et al. 2001a), SO, CH$_3$OH (Lis et al. 2001) are measured towards large regions ($>10$ pc) in the whole GC. These molecules are usually in the grain mantles. The relatively low dust temperatures in the GC and the extended emission of those molecules imply that the processes giving rise to the measured high gas-phase abundances are different than those taking place in a hot-core, where they are explained by thermal evaporation from dust grains with a temperature higher than 70 K. Furthermore, as discussed above, these molecules are easily dissociated in the presence of UV radiation. Therefore, their high gas phase abundances and their extended distribution suggest that they have been ejected from the grains by mechanical processes, such as low velocity shocks. These have also been invoked to explain the transitory gas heating and the non-equilibrium H$_2$ ortho-to-para ratio measured in two GC clouds (Rodríguez-Fernández et al. 2000).

The mechanical processes could be  turbulence or  shocks. Rodriguez-Fernandez et al. (2001a) showed that the energy in turbulent motions is enough to explain the cooling of the 150 K gas by CO and H$_2$ rotational lines. It is quite difficult to deal with the turbulence. Current models do not make direct predictions of lines intensities but explain the processes taking place in the turbulent interstellar medium. For instance, Smith et al. (2000) have shown how the turbulence creates a full spectrum of shocks with different velocities. They found that the number of shocks is inversely proportional to the square of the shock velocity. Hence, most of the gas mass is suffering slightly supersonic shocks with low Mach numbers of 1-5, which for the GC warm gas implies very low shock velocities of 1-5 \kms. Indeed, the velocity of the shocks taking place in the GC must be low enough to heat the H$_2$, but without giving rise to much fine-structure line emission since, as we have already shown, the observed fine-structure emission can be accounted for by the PDR component alone. The key in the GC could be the magnetic field strength, which can be as high as 1 mG. In the presence of a strong field C-shocks have an important magnetic precursor that heats large quantities of gas to moderate temperatures ($\sim$ 150 K). For instance, a  C-shock with velocity of 25 \kms ~ in a preshock gas density of 10$^4$ \cmmt ~ can reproduce the S(0) and S(1) H$_2$ intensities observed in the GC without giving much emission in fine structure lines ([\OI] and [\CII] lines intensities lower than $10^{-6}$ erg\,\smu\,\cmmd\,sr$^{-1}$; S. Cabrit priv. communication).

The origin of the  turbulence and shocks in the GC can be local or related to the large-scale properties of the Galaxy. On the one hand, massive stars are an important source of shocks and turbulence in scales of a few pc by means of supernova blast waves, \HII ~ regions and strong winds. For instance, hot (60-130 K) NH$_3$ shells with sizes of 1-2.5 pc have been observed in the envelope of the Sgr B2 molecular cloud (Martín-Pintado et al. 1999). These shells expand  with a low velocity of $\sim 5$ \kms ~ and are probably created by the interaction of Wolf-Rayet stars winds with the surrounding ISM. The scenario found in the Sgr B2 envelope would explain the non-homogeneity of the GC ISM required to explain the gas ionization (Rodríguez-Fernández et al. 2001b, Rodríguez-Fernández and Martín-Pintado 2004). In addition  these findings have important implications in the heating of the GC molecular gas. The UV radiation from the stars responsible for the ionization will also create PDRs in the edges of the cavities. In addition,  C-shocks with a velocity similar to the shell expansion are also expected.

On the other hand, the origin of the turbulence can be tidal forces induced by the rotation of the Galaxy, as proposed by Wilson et al. (1982) to account for the high temperatures of  the molecular gas in the Sgr B2 envelope. Indeed, the models by Das \& Jog (1995) have shown that tidal forces can  heat the molecular clouds moving along elongated orbits in a bar potential, which seems to be the case in the GC (Binney et al. 1991).

The GC is a complex region, where many processes can play a role in the excitation of the ISM. We conclude that,  extended PDRs  can explain the fine structure line emission and the H$_2$ emission from excited levels. On the other hand, the most likely heating source of the bulk of 150 K molecular gas, traced by the H$_2$ emission from low levels (and the GC chemistry) are moderate velocity shocks  most likely induced by the turbulent nature of the GC ISM.

%.......................................................
\section{Conclusions}

We have presented fine-structure line data emitted by the main coolants of the neutral gas and the dust continuum spectra around the emission peak. The spectra were taken towards  a sample of molecular clouds distributed along the 500 central pc of the Galaxy and located far from thermal radio continuum or FIR sources. We also compared these results with  the H$_2$ pure-rotational lines presented by Rodríguez-Fernández et al. (2000, 2001a).

The dust continuum emission can be modeled with two temperatures: a cold component with a temperature of $\sim 15$ K and a warmer component whose temperature varies between 25 to 40 K from source to source. 

We have compared the fine-structure lines and the dust continuum emission with the predictions from shocks and PDR models. We have  found that the observations are best explained by a PDR with a density of 10$^3$ \cmmt ~ and a far-ultraviolet incident field 10$^3$ times higher than in the local ISM. The PDRs should be located in the interfaces between the dense molecular matter and the ionized gas component discussed by Rodríguez-Fernández and Martín-Pintado (2004). Indeed, the [\CII] 158 \mum ~ and the [\SiII] 34 \mum ~ lines have a significant contribution from the ionized gas component.

The H$_2$ pure-rotational emission from excited levels (Rodríguez-Fernández et al. 2000, 2001a) and the ro-vibrational near-infrared emission (Pak et al. 1996) also arise in the PDR component. However, only 10-30 $\%$ of the gas traced by the lowest pure-rotational levels (S(0) and S(1)), with a temperature of $\sim 150$ K, can be accounted for in this context. We suggest that the most promising heating mechanism of this gas (and the origin of the  chemistry characteristic of the GC) are low velocity C-shocks created by turbulent motions.

Higher spatial and spectral resolution  as well as mapping observations  of the warm H$_2$ and key fine-structure lines from neutral and ionized atoms are needed to get further insight in the interplay of the different ISM phases and the heating of the molecular gas in the GC. The new generation of medium and far-infrared telescopes (VLT, SOFIA, Herschel, ALMA) is well suited to attack these problems in the GC and other nearby galaxies.

\begin{acknowledgements}
NJR-F acknowledges Mario van den Ancker for providing us with the shocks and PDR model data in digital form to produce Fig. 7. NJR-F also acknowledges fruitful discussions with J.R. Goicoechea (Sgr B2), M. Morris (heating and ionization) and S. Cabrit (C-shocks). NJR-F has been supported by a Marie Curie  Fellowship of the European Community program ``Improving Human Research Potential and the Socio-economic Knowledge base'' under contract number HPMF-CT-2002-01677.  JM-P has been partially supported by the Spanish {\it Ministerio de Ciencia y Tecnologia} (MCyT) grants AYA2002-10113-E, AYA2003-02785-E and ESP2002-01627. AF has been funded by the Spanish McyT grants DGES/AYA2003-07584 and ESP2003-04957. This research has made use of data products from the Midcourse Space Experiment. Processing of the data was funded by the Ballistic Missile Defense Organization with additional support from NASA Office of Space Science. The data were accessed by services provided by the NASA/IPAC Infrared Science Archive.
\end{acknowledgements}

\end{document}